# Intelligent Momentary Assisted Control for Autonomous Emergency Braking

Konstantinos Gounis[1], Nick Bassiliades[1*]

[1]Department of Informatics, Aristotle University of Thessaloniki, 54124 Thessaloniki, Greece

*Corresponding author email: nbassili@csd.auth.gr

## Abstract

Development of control algorithms for enhancing performance in safety-critical systems such as the Autonomous Emergency Braking system (AEB) is an important issue in the emerging field of automated electric vehicles. In this study, we design a safety distance-based hierarchical AEB control system constituted of a high-level Rule-Based Supervisory control module, an intermediate-level switching algorithm and a low-level control module. The Rule Based supervisor determines the required deceleration command that is fed to the low-level control module via the switching algorithm. In the low-level, two wheel slip control algorithms were developed, a Robust Sliding Mode controller and a Gain-Scheduled Linear Quadratic Regulator. For the needs of this control design, a non-linear dynamic vehicle model was implemented whereas a constant tire-road friction coefficient was considered. The proposed control system was validated in Simulink, assuming a straight-line braking maneuver on a flat dry road. The simulation results demonstrated satisfactory emergency braking performance with full collision avoidance in both proposed control system combinations.

1.Introduction

Road accidents account for a considerable loss of lives worldwide. In the European Union, 25,000 deaths are occurring every year due to road accidents, according to EU statistics [1]. In order for this problem to be eliminated, new mandatory safety systems have been agreed to be part of the passenger vehicles' equipment as of 2022, such as Lane Keeping Assistance, Advanced/Autonomous Emergency Braking, Intelligent Speed Assistance, warning systems, and crash-test improved safety belts. It is estimated that the proposed measures will help to save over 25,000 lives and avoid at least 140,000 serious injuries by 2038 [1]. These goals comply with Vision Zero, the EU's long-term goal of moving close to zero fatalities and serious injuries by 2050. Road fatalities is also a problem outside Europe. In China, for example, 63,700 people lost their lives during 2017 [2]. In addition to this, car collision accidents account for about 70%, the majority of which are rear-end accidents[2]. In the EU, it has been reported that the 39% of fatal road accidents are occurring in urban/inter-urban roads [3]. As car to car 'nose-to-tail' collisions are a common type amongst them, Advanced Emergency Braking is of particular importance for collision avoidance.

According to EU legislation, Advanced Emergency Braking is defined as the system comprised of exteroceptive sensor(-s) and the control module which can automatically detect a potential collision and activate the vehicle braking system to decelerate the vehicle with the purpose of avoiding or mitigating a collision [4]. The terms "Advanced", "Automatic" and "Autonomous" in the context of emergency braking for low-level vehicle autonomy seem to be used interchangeably [5,6]. Regulation No 131 of the Economic Commission for Europe of the United Nations (UN/ECE) states that the emergency braking phase is the interval starting when the AEB system emits a braking demand for at least 4 $m/s^2$ deceleration to the service braking system of the vehicle [7]. Society of Automotive Engineers (SAE) organization, in SAE J3016 (Levels of Vehicle Automation), defines Automatic Emergency Braking as a function that is limited to providing warnings and momentary assistance to a modern car [5].

In the previous years, automated driving has been scientifically approached with planning and control methods that were developed upon several assumptions, such as: a) steady state, low speed operation, b) no slipping / no sliding conditions, and c) linearized dynamics, away from the adhesion limits [8]. Such assumptions make the vehicular motion control valid in a range of dynamics only, corresponding to the way average drivers operate their automobiles. *Velenis* [8] through his comprehensive research in longitudinal and lateral motion control, developed methods and showed that the vehicle can be controlled directly via the longitudinal slip at each wheel.

Euro NCAP in the 2018 test protocols concerning car-to-car rear braking (CCRb), car-to-car rear stopping (CCRs) and car-to-car rear moving (CCRm), defines time to collision (TTC) as the remaining time before the vehicle equipped with AEB strikes the leading vehicle, assuming that both vehicles would continue to travel with the speed they are travelling at the moment TTC was calculated [6]. In the study of *Das et al.* [9], the same TTC definition is presented. However, TTC metric adoption inherits some modeling risks. This is due to its mathematical definition, being the ratio of relative distance divided by relative speed, which might lead to undesired calculation results. This issue is addressed by introducing a more complex piecewise TTC formula or by using the safety distance metric. Recent research work on AEB tends to adopt distance-based metrics [10].

Although the study of Autonomous Vehicle Control has been a commonplace [11-15], less studies have targeted control methods/algorithms to avoid or mitigate collision in the context of AEB. Most of these studies have considered the Time-to-Collision metric. *Han et al.* [16] proposed a TTC-based AEB control system considering the varying road-tire friction. *Shin et al.* [17] demonstrated an adaptive TTC-based AEB control strategy considering both the threat that occurs at the front of the vehicle under consideration (e.g. pedestrian, leading vehicle) and a possible collision risk with a vehicle on the rear of the vehicle under consideration. *Guo et al.* [18] focused on a Variable Time Headway-based safety distance model, to address collision avoidance by using Model Predictive Control (MPC) in a system governed by linear vehicle dynamics. *Yang et al.* [2] established an Autonomous Emergency Braking Pedestrian (AEB-P) warning model, considering not only TTC but also braking safety distance. This model incorporated a Neuro-Fuzzy system trained on collected anti-collision braking operation data of experienced drivers.

*Kim et al.* [10], on the other hand, developed an AEB control algorithm which is purely distance based. In their study, the Minimum Stopping Distance metric was defined. Based on a few different car-to-car-rear braking, moving and stopping driving scenarios, discrete minimum braking and stopping distance formulas for each driving scenario were defined. A desired deceleration command was calculated, based on the aforementioned formulas, which was then regulated to a low-level PI controller that delivered the deceleration requests. A point mass longitudinal dynamics vehicle model was considered, taking into account road slope and friction coefficient, the latter in the context of linear longitudinal-normal load relationship.

Emergency braking has also been approached from the-road friction estimation necessity-point of view. *Alvarez et al.* [19] focused their study on the road tire coefficient $\mu$ estimation via state observers and the asymptotic stability of the control system they proposed. Emergency deceleration is highly influenced by wheel slip control whereas a critical parameter for autonomous vehicle dynamics control is the longitudinal slip [8]. Wheel slip control design has been recently reviewed comprehensively by *Pretagostini et al*. [20]. In this review, the prevailing wheel slip control methods, namely Rule Based control, Fuzzy Based control, PID control, Sliding Mode control (SMC), Robust control, Neural Network based control, Linear Quadratic Regulator (LQR) based control and Model Predictive Control were evaluated. It was shown that SMC and LQR are amongst the best wheel slip control strategies demonstrating high setpoint tracking capability, relatively high robustness and adaptability and moderate computational intensity.

Taking together the recent EU requirements towards Vision Zero and the shortage of publications on distance-based, non-linear control approaches for AEB design, we propose an intelligent momentary assisted control for AEB. In the proposed AEB control: a) the collision risk is determined in terms of the relative distance to the leading vehicle which is compared against an adaptive velocity based relative distance threshold, b) the outcome of this comparison is fed to a Rule Based Supervisor, and c) the low-level control module has been designed so that it efficiently regulates the longitudinal slip target that corresponds to the desired deceleration. Focusing on robust and adaptive wheel slip control, a Sliding Model control algorithm and a Gain Scheduled Linear Quadratic Regulator were designed and compared in the context of the AEB performance.

## 2. Modeling and Control

In this section, the modeling of the system as well as the Autonomous Emergency Braking (AEB) control design are presented. The dynamical model representing the vehicle equipped with AEB (EGO vehicle hereinafter) is comprised of a system of non-linear differential equations. These equations will be referred hereinafter as *longitudinal and lateral equations of motion*. As the maneuver under consideration is an autonomous straight-line braking, the equations of motion are simplified to describe non-linear, pure longitudinal motion. The dynamics that represent the leading vehicle's motion are not analyzed. Instead, given the leading vehicle's trajectory, the non-linear dynamics of the EGO vehicle are analyzed and the AEB control system is designed so that it delivers the required emergency braking maneuver. The AEB control system follows a hierarchical structure. More specifically, the high-level control is provided by the Rule Based Supervisory Control system that plans the deceleration actions. The intermediate-level control is provided by the Switching Algorithm that coordinates the control actions based on the severity of the potential collision risk. The low level-control delivers the commands from the upper levels by robustly controlling the slip dynamics. The environment is comprised of a flat, dry road. Figure 1 demonstrates the schematic of the EGO-leading vehicle interaction. Figure 2 depicts in more detail the architecture of the EGO vehicle with the hierarchical control structure.

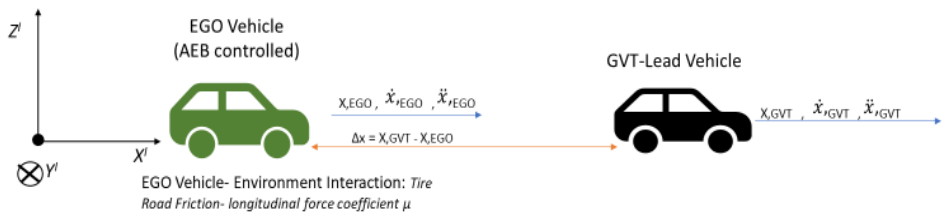

*Figure 1. Illustration of the EGO-Lead Vehicle interaction*

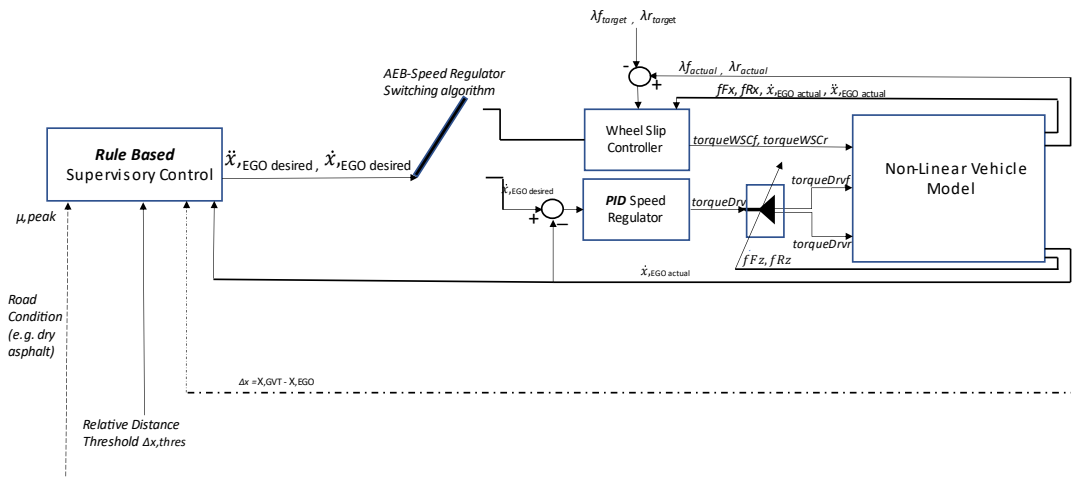

*Figure 2. Illustration of the EGO vehicle, supplied with the hierarchical-structured AEB control system, where $\lambda f_{target}$, $\lambda r_{target}$ are front and rear longitudinal slip targets.*

As wheel slip control is of particular importance in Autonomous Emergency Braking applications, two different low-level controllers are designed, from the control algorithm perspective: a) a Robust Sliding Mode wheel slip controller and b) a Gain-Scheduled Linear Quadratic Regulator for the emergency braking maneuver stabilization. These controllers are analyzed in the related sections.

*2.1 EGO Vehicle Modeling*

In this study, dynamics of the EGO vehicle are represented by the dynamic non-linear bicycle model coupled with non-linear tire models. This single-track model achieves a satisfactory level of fidelity [21] as it can be considered an equilibrium point on the tradeoff between high fidelity, computationally intense models (e.g. high order Multi-Body-Dynamics Models) and low fidelity, computationally cheap models (e.g. Point Mass models, kinematic models with linear tire response, etc.). Constructing this model, the following assumptions have been made:

   a) On each axle, the two wheels are represented by a single equivalent wheel with characteristics that will be presented later in this section.
   b) Total vehicle mass is lumped in the vehicle center-of-gravity (C.O.G) whereas reaction (normal) loads are occurring on the front and rear axles, respectively.
   c) Tire longitudinal/lateral friction forces on each axle are estimated using the Pacejka MF model.
   d) This vehicle architecture can encompass electric prime movers on each axle, all of them having the ability to serve as regenerative braking actuators.
   e) It is assumed that required braking torque is provided by the best strategy, either hydraulic friction brake (HFB) assisted or electric motor and HFB assisted.
   f) The braking torque blending strategy, as well as the modelling of the electric motors and the hydraulic brake system, are out of the scope of this study.
   g) Brake actuation delays are neglected.

The vehicle's equation of motion is:

$$\ddot{x}_{,EGO\ actual} = \frac{fFx + fRx}{m_{,veh}} \quad (2.1.1)$$

where $\ddot{x}_{,EGO\ actual}$ is the acceleration component along the x axis of the inertial frame of reference $\{X^I Y^I Z^I\}$ fixed at the origin, $fix$ are the longitudinal forces of the front/rear tires ($i=\{F, R\}$) and $m_{,veh}$ is the total vehicle mass.

In this study, the practical slip ratio definition [22] is used for the needs of the wheel slip control design and control logic deployment:

$$\lambda i = \frac{Vix - \omega_i R}{|Vix|} \quad (2.1.2)$$

where $\omega_i$ is the rotational speed of the front/rear wheel ($i=\{F, R\}$) and $Vix$ is the translational speed of the front/rear wheel.

The longitudinal friction coefficients $\mu iX$, estimated using Pacejka's Magic Formula model [22] are given by:

$$\mu iX = D sin(C tan^{-1}(BsiX)) \quad (2.1.3)$$

where $D$ is the coefficient that represents peak road-tire adhesion, $B$ is the stiffness coefficient, $C$ is the shape factor (Table 1) and $siX$ is the theoretical front/rear slip ratio [22].

The normal load on each tire accounts for the static load at zero acceleration/deceleration plus the dynamic load due to longitudinal load transfer, defined as follows [21]:

$$fFz = \frac{lr\, m_{,veh}\, g - h\, m\, g\, \mu Rx}{lf + lr + h(\mu Fx - \mu Rx)} \quad (2.1.4)$$

$$fRz = m_{,veh}\, g - fFz \quad (2.1.5)$$

where $g$ is the gravitational acceleration, $lf, lr$ are the distances of the front and rear axle from the vehicle's center-of-gravity (C.O.G) and $h$ is the height of the C.O.G. The longitudinal friction forces are equal to:

$$fix = \mu iX \times fiz, \; i=\{F, R\} \quad (2.1.6)$$

Finally, the wheel dynamics are presented, as they are of particular importance, in the context of Autonomous Emergency Braking control design:

$$\dot{\omega}_{front} = (T_{front} - fFxR)/I \quad (2.1.7)$$

$$\dot{\omega}_{rear} = (T_{rear} - fRxR)/I \quad (2.1.8)$$

where $\dot{\omega}_{front}$, $\dot{\omega}_{rear}$ are the front and rear wheel rotational accelerations and $I$ is the wheels' mass moment of inertia. $T_{front}$, $T_{rear}$ are the front and rear control torques that will be provided either by the wheel slip control module when there is urge for momentary emergency braking assistance or by the speed regulator in case of cruising operation. These control modules as well as the high-level supervisory control along with the switching algorithm to intervene between emergency and non-emergency situations, will be analytically described in the next section.

Table 1. Vehicle Data used in this study, corresponding to the electric vehicle used in Siampis et al.[23]

| | |
|---|---|
| $m_{,veh} = 1420\, kg$ | $h = 0.55\, m$ |
| $g = 9.81\, m/s^2$ | $R = 0.3\, m$ |
| $B = 24, C = 1.5, D = 0.9$ | $Iz = 1027.8\, kgm^2$ |
| $lf = 1.01\, m, lr = 1.452\, m$ | $I = 0.6\, kgm^2$ |

## 2.2 Autonomous Emergency Braking Control

*2.2.1.1 Rule based supervisory control* We consider a high-level control system driven by Rule Based logic in order to ensure that a target deceleration defined by the tire's maximum capacity at the road friction limit is provided. The Rule Based Supervisory Control (RBSC hereinafter) supervises the AEB system's progress towards achieving its goal by devising a set of verbal rules. Rule Based systems include a Rule Base coupled to an Inference Mechanism. These systems' typical operation implies that given some input data, the system will be capable of drawing meaningful conclusions, according to the conditions that the system meets [24]. In the context of this study, the conditions the vehicle under test (EGO hereinafter) meets are considered and based on the execution of the rules a desired output to follow is generated. The inputs to the RBSC are: peak friction coefficient based on road condition $\mu$, EGO vehicle's actual longitudinal speed $\dot{x}$,EGO actual, relative distance of lead vehicle (GVT hereinafter) w.r.t EGO vehicle $\Delta x$, and a defined relative distance threshold $\Delta x, Thres$, which acts as an adaptive tuning parameter in order to establish a safety distance from the GVT at all times. In this study, it is assumed that the GVT is precisely tracked by the exteroceptive sensors (e.g. Radar, Lidar) and that the EGO vehicle's speed is available by internal sensors, State Estimators and/or GPS at all times.

Peak achievable deceleration is defined as $-\mu \ast g$, where $g$ is gravitational acceleration.
EGO vehicle's minimum braking distance $X_{br,min}$ is defined as follows:

$$X_{br,min} = \frac{(\dot{x},\text{EGO actual})^2}{2 \ast \mu \ast g} \qquad (2.2.1.1.1)$$

where $\dot{x},_{\text{EGO,actual}}$ is the EGO vehicle's longitudinal speed at a time instance.

Based on *(2.2.1.1.1)*, the relative distance threshold is a function of speed, plus an additional static safety distance margin (*margin*):

$$\Delta x, Thres = X_{br,min} + margin \qquad (2.2.1.1.2)$$

It is obvious that relative distance threshold is capable of adaptation to the velocity the vehicle undergoes at a time instance.

Finally, the target deceleration is defined as:

$$\ddot{x}_{\text{EGO,desired}} = \frac{(\dot{x},\text{EGO actual})^2}{-2 \ast xbr,min} \qquad (2.2.1.1.3)$$

Taking into consideration equations *(2.2.1.1.1)- (2.2.1.1.3)*, the RBSC logic is assembled as follows:

```
ON    DrivingwithLeadingTraffic

IF    DeltaX is equal to or less than DeltaXthres

THEN

      applydecel(Desiredecel=TARGET)

ELSE

      applydecel(Desiredecel=ZERO)
```

As stated above, the logic consists of an Event-Driven Rule [25] that is activated when longitudinal motion is detected on the EGO vehicle and a braking/slow moving leading vehicle is detected. In addition to this, the Reasoning method RBSC adheres to is forward chaining.

RBSC logic can be readily transformed to a non-symbolic format that consists of the following algorithms:

**Algorithm 1: *targetXdotdotGenActivation***

*Input: $\dot{x},_{\text{EGO,actual}}$, bool isLeadingVehicleDetected, $\mu_{peak}$*

**while** *( $\dot{x},_{\text{EGO,actual}}$ >4) && ( isLeadingVehicleDetected == 1)* **do**

  *targetDecelerationGenerator($\Delta x$, $\dot{x},_{\text{EGO,actual}}$, $\mu_{peak}$)*

**endwhile**

**end**

## Algorithm 2: targetDecelerationGenerator

*Input:* $\Delta x$, $\dot{x}_{EGO,actual}$, $\mu_{peak}$

*Output:* $\ddot{x}_{EGO,desired}$

#define g 9.81

#define margin m   //additional safety distance margin

$\ddot{x}_{EGO,desired} \leftarrow 0$

$x_{br,min} \leftarrow -(\dot{x}_{EGO,actual}) * (\dot{x}_{EGO,actual}) / (2*(-\mu_{peak})*g)$

$\Delta x_{thres} \leftarrow x_{br,min} + m$

**if** $(\Delta x <= \Delta x_{thres})$

    $\ddot{x}_{EGO,desired} \leftarrow - (\dot{x}_{EGO,actual}) * (\dot{x}_{EGO,actual}) / (2*x_{br,min})$

**else**

    $\ddot{x}_{EGO,desired} \leftarrow 0$

**endif**

*switchingAlgorithm($\ddot{x}_{EGO,desired}$, $\dot{x}_{EGO,actual}$)*

**return** $\ddot{x}_{EGO,desired}$

## Algorithm 3: switchingAlgorithm

*Input:* $\ddot{x}_{EGO,desired}$, $\dot{x}_{EGO,actual}$

*torqueWSCf* $\leftarrow 0$

*torqueWSCr* $\leftarrow 0$

*torquedrv* $\leftarrow 0$

**if** *abs($\ddot{x}_{EGO,desired}$)>0*

   *torqueWSCf* $\leftarrow$ *SlidingModeControl(isActive=1)*

   *torqueWSCr* $\leftarrow$ *SlidingModeControl(isActive=1)*

   *torquedrv* $\leftarrow$ *ProportionalIntegralDerivativeControl(isActive=0, $\dot{x}_{EGO,actual}$, $\ddot{x}_{EGO,desired}$)*

**else**

   *torqueWSCf* $\leftarrow$ *SlidingModeControl(isActive=0)*

   *torqueWSCr* $\leftarrow$ *SlidingModeControl(isActive=0)*

   *torquedrv* $\leftarrow$ *ProportionalIntegralDerivativeControl(isActive=1, $\dot{x}_{EGO,actual}$, $\ddot{x}_{EGO,desired}$)*

**endif**

**end**

*2.2.1.2 PID Speed Regulator*

PID controllers are typical elements in automated vehicles as they can provide sufficient low-level control such as steering/braking/traction commands [26]. In this study, we consider a PID Controller that regulates EGO vehicle speed when there is no request by the RBSC for Autonomous Emergency Braking (AEB) deceleration. As the acceleration target is 0, the desired speed is defined as the speed EGO vehicle should cruise at when the distance between GVT and EGO just exceeded the minimum relative distance threshold $\Delta x, Thres$ ; hence, the situation is no longer considered worrying. The latter improves the overall motion control in the following manners: a) EGO vehicle manages to maintain a constant (cruising) speed by balancing the non-linear friction forces exerted at the wheels and b) an additional control logic has been established for the in-between threat overcoming phases. The leading vehicle may further brake afterwards; hence, the AEB cycle will be repeated if required.

The desired longitudinal speed is defined as follows:

$$\dot{x}_{,EGO,desired}(t) = \int_0^t \ddot{x}{EGO, desired}\, dt' + v(0) \qquad (2.2.1.2.1)$$

The error $e(t)$ is defined as the difference between desired and actual speed:

$$e(t) = \dot{x}_{,EGO,desired}(t) - \dot{x}_{,EGO,actual}(t) \qquad (2.2.1.2.2)$$

The proposed feedback controller is:

$$\boldsymbol{torquedrv(t)} = \boldsymbol{K} * sat(kp * e(t) + ki \int_0^t e(t)dt' + kd * \frac{de(t)}{dt} * \frac{N}{1+N \int_0^t dt'}) \qquad (2.2.1.2.3)$$

where:

$sat(.)$ is a function that saturates proportional-integral-derivative control actions considering braking torque limitations as well as adhesion limitation, $N$ is a filter coefficient applied to the derivative part, $kp, ki, kd$ are the selected PID gains guaranteeing the error polynomial is Hurwitz and $K$ is a vector valued adaptive gain that comprises of non-linear elements:

$$K = \left[\frac{fFz(\mu Fx, \mu Rx, \mu Fy)}{fFz + fRz}, \frac{fRz(\mu Fx, \mu Rx, \mu Fy)}{fFz + fRz}\right]^T \qquad (2.2.1.2.4)$$

whilst $\boldsymbol{torquedrv(t)}$ is constituted of the front and rear axle torque components:

$$\boldsymbol{torquedrv(t)} = [torquedrvf(t), torquedrvr(t)]^T \qquad (2.2.1.2.5)$$

Therefore, the torques are provided to the front/rear axle by devising a torque allocation based on adhesion utilization. A similar approach is proposed in Ruiz Diez et al. [27].

*2.2.2 Sliding Mode wheel slip control*

Sliding Mode Control is a non-linear control method that aims at driving the states of variable structure real-world systems towards their desired values. This is achieved by applying control actions that alter the dynamics of the systems in a fashion of applying discontinuous signals. The aforementioned signals exhibit a switching behavior in-between some mathematically defined control boundaries [28]. In the application field of wheel slip control, Sliding Mode is often selected as the basic control logic due to its robustness in the presence of parameter variations and disturbances [29, 30].

We consider the system:

$$x^{(n)}(t) = f(\mathbf{x},t)+g(\mathbf{x},t) \qquad (2.2.2.1)$$

where $x^{(n)}(t) = d^n x(t)/dt^n$, $\mathbf{x}(t) =[x, \dot{x}, \ldots, x^{n-1}]^T$ is the state vector, $f(\mathbf{x},t)$ and $g(\mathbf{x},t)$ are non-linear functions of time and state. The rationale is to define a control law $u=g(\mathbf{x},t)$ that guarantees that the tracking error vector $\mathbf{e}(t)= \mathbf{x}(t)-\mathbf{x}_{des}(t)= [e, \dot{e}, \ldots, e^{n-1}]^T$ approaches zero asymptotically ($e(t) \to 0$ for $t \to \infty$).

The sliding surface $s(x,t)$ is defined as:

$$s(x,t) = \left(\frac{d}{dt} + \xi\right)^{n-1} e(t) = \sum_{k=0}^{n-1}\binom{n-1}{k}\xi^{\kappa} e^{n-1-k} \quad , e(0) \text{ known} \qquad (2.2.2.2)$$

By setting $s(x,t) = 0$, equation (2.2.2.2) becomes a homogenous differential equation w.r.t tracking error that has a unique solution for each $t$ and also satisfies the given initial condition.

Defining practical wheel longitudinal slip $\lambda i$ as the system state and the tracking error $e(t)$ as $e(t)=\lambda i(t)-\lambda i_{ref}(t)$, i={F, R}, equations (2.2.2.1) and (2.2.2.2) are converted to:

$$\frac{d\lambda i}{dt} = -\frac{R}{J\,Vix}(Ti - fixR) + \frac{\dot{V}ix}{Vix}(1 - \lambda i) \qquad (2.2.2.3)$$

$$s(\lambda, t) = 0, \ s = \lambda i - \lambda i, ref, \ \dot{s} = \dot{\lambda}i - \dot{\lambda}i, ref \qquad (2.2.2.4)$$

Aiming for a steady state, controlled slip condition ($\dot{\lambda}i, ref = 0$), one gets:

$$\dot{s} = \dot{\lambda}i \qquad (2.2.2.5)$$

Parameter $\xi$ in (2.2.2.2) gives quantitative and qualitative knowledge about the cut-off of the system's un-modelled frequencies according to Tzafestas [31]. In this study, $\xi$ is readily eliminated.

Taking into consideration all the aforementioned, in order to derive a control law $u=g(x,t)$ that guarantees $e(t) \to 0$ for $t \to \infty$, a Lyapunov function is defined as:

$$V = \frac{1}{2}s(x,t)^2 \ , \ V(0)=0, \qquad (2.2.2.6)$$

that satisfies the stability condition:

$$\dot{V} = \frac{1}{2}\frac{d}{dt}[s(x,t)^2] \leq -\eta|s| \qquad (2.2.2.7)$$

Equation (2.2.2.7) leads to the following:

$$s\dot{s} \leq -\eta|s| \implies \dot{s}\,sign(s) \leq -\eta \qquad (2.2.2.8)$$

As long as $\eta \geq 0$, the system is operating in sliding mode. The objective is to derive a control signal that satisfies the stability condition. Replacing (2.2.2.5) in (2.2.2.8) gives:

$$s\,\dot{\lambda}i \leq -\eta\,|s| \qquad (2.2.2.9)$$

Equation (2.2.2.3) combined with (2.2.2.9) result in:

$$s\left[-\frac{R}{JVix}(Ti - fixR) + \frac{\dot{V}ix}{Vix}(1 - \lambda i)\right] \leq -\eta|s| \qquad (2.2.2.10)$$

The control law used in this study is:

$$Ti = T_{eq} + T_s\, sign(s) \qquad (2.2.2.11)$$

where $T_{eq}$, referred as equivalent control [32], is the required control in order for the system dynamics to perform an ideal sliding motion on the surface, obtained by setting $\dot{\lambda}i = \dot{\lambda}i, ref = 0$ and solving equation *(2.2.2.3)*. $T_s\, sign(s)$, on the other hand, is the switching torque that enforces the error dynamics to return to the sliding surface $s(\lambda, t) = 0$, i.e. "pressing" the system dynamics whenever it escapes from the surface to return to its controlled sliding motion.

Replacing $Ti$ with $T_{eq} + T_s\, sign(s)$ and solving the inequality *(2.2.2.10)* lead to the following piecewise function for switching torque:

$$Ts = \begin{cases} \dfrac{-\eta J Vix}{R} & if\ (\lambda i - \lambda i, ref) \leq -1, \\ \dfrac{\eta J Vix}{R} a\, sign(\lambda i - \lambda i, ref)|\lambda i - \lambda i, ref| & if\ 0 \leq |\lambda i - \lambda i, ref| \leq 1, \\ \dfrac{\eta J Vix}{R} & if\ (\lambda i - \lambda i, ref) \geq 1, \end{cases} \qquad (2.2.2.12)$$

Parameters $a$ and $\eta$ represent the slope in the second part of the non-linear piecewise function and the positive number for the stability condition in *(2.2.2.8)* to hold, respectively. The aforementioned parameters were tuned, considering the need for chattering reduction of the error signal in the sliding manifold, and thus a robust wheel slip controller was implemented for the AEB system.

The control law in *(2.2.2.11)* is equivalent to the following Reinforcement Learning based control algorithm that defines a deterministic mechanism that optimizes policy $\pi$ ($\pi$ equivalent to control law *u*), based in Lyapunov Stability theory [33]:

**Algorithm 4: PolicyIterationBasedRobustControlAlgorithm**

*Input:* $e(0)$, *π(0)  //Initial Conditions*

*Output: $\pi^*$ $V^*$ // outputs optimal policy $\pi^*$ and the cost function V value that has been minimized cost-wise (sign can //be reverted to match reward maximization problem): V(t) = $e(t) = -e(t-1)$, $e(0)\ known$ //, in ODE format written as $\dot{e} = -e$, $e(0)\ known$*

*#define K //K is a gain derived from the Lyapunov Stability condition in eq.2.2.2.8*

**repeat** *at every iteration t=1,2,…,tfinal*

   **find** *$V^\pi$ //evaluate the cost function value that policy π has given so far*

  *$\pi_0$ (t) ← f(($\dot{x}$=0),t-1)  //equivalent control action as per eq.2.2.2.10, where f(.) is a non-linear process*

  *π(t) = K\*sign(e(t-1))\*|e(t-1)| +π₀(t)*

**until** *π(t) =π(t-1)*

**return** *$\pi^*$, $V^*$*

## 2.2.3 Trajectory stabilization for the emergency braking condition via Gain-Scheduled LQR control

In order to compare Sliding Mode Wheel Slip Control based emergency braking with another State-of-the-art wheel slip control method, a linear quadratic regulator (LQR) is designed to control the non-linear vehicle model during the emergency maneuver. As the system cannot be expressed as a Linear Time Invariant state space model, the gain scheduling approach is adopted to tackle challenges that non-linearity inherits.

The wheel dynamics comprise of a non-linear system of the form:

$$\dot{x} = \begin{bmatrix} f(x,u) \\ g(x,u) \end{bmatrix}$$

$$y = h(x) \quad (2.2.3.1)$$

Setting $\dot{x} = \begin{bmatrix} \dot{V}ix \\ \dot{\omega}i \end{bmatrix}$, $u = Ti$ and $y = \lambda i$, where $Vix$ is the longitudinal speed of the {front, rear} wheel and $\omega$ the rotational speed of the {front, rear} axle, we obtain:

$$\begin{bmatrix} \dot{V}ix \\ \dot{\omega}i \end{bmatrix} = \begin{bmatrix} -gD\sin(C\tan^{-1}(B\lambda i)) \\ \left(\frac{1}{J}\right) mgD\sin(C\tan^{-1}(B\lambda i))R + \left(\frac{1}{J}\right)Ti \end{bmatrix} \quad (2.2.3.2)$$

where $m$ corresponds to front/rear axle load divided by gravitational acceleration. As previously stated, longitudinal load transfer is considered too, in this study, as it is strongly influencing tire longitudinal friction forces.

Emergency deceleration, thus, can be defined as a transient reference trajectory with the following dynamics, based on a constant slip target $\lambda, ref$ that corresponds to the required vehicle deceleration for the given road condition:

$$\dot{V}xi, ref = -gD\sin(C\tan^{-1}(B\lambda i, ref)) \quad (2.2.3.3)$$

$$\omega i, ref = Vxi, ref(1 - \lambda i, ref) \quad (2.2.3.4)$$

$$Ti, ref = J\dot{\omega}i, ref - (m \pm \Delta mi, ref)gD\sin(C\tan^{-1}(B\lambda i, ref))R \,, \; m \pm \Delta mi, ref =$$

$$\frac{m_{veh}li}{lf+lr} \pm \frac{m_{veh}\left(\frac{\dot{V}xi,ref}{g}\right)h}{lf+lr} \,, i = \{F, R\} \quad (2.2.3.5)$$

Equation *(2.2.3.4)* forms the reference control input. In order to ensure that EGO vehicle stabilizes around the required reference trajectory, the system described in *(2.2.3.2)* was linearized and afterwards feedback controlled about the discussed reference trajectory. The Jacobians of the system were calculated to obtain the incremental model. Ignoring high-order terms, one gets:

$$\begin{bmatrix} \Delta\dot{V}xi \\ \Delta\dot{\omega}i \end{bmatrix} = \begin{bmatrix} \frac{\partial f}{\partial Vxi} & \frac{\partial f}{\partial \omega i} \\ \frac{\partial g}{\partial Vxi} & \frac{\partial g}{\partial \omega i} \end{bmatrix} \Big|_{Vxi=Vxi,ref \; \omega i=\omega i,ref} \begin{bmatrix} \Delta Vxi \\ \Delta \omega i \end{bmatrix} + \begin{bmatrix} \frac{\partial f}{\partial Ti} \\ \frac{\partial g}{\partial Ti} \end{bmatrix} \Big|_{Ti=Ti,ref} \Delta T \,, \; \Delta Ti = Ti -$$

$Ti, ref, \Delta Vxi = Vxi - Vxi, ref, \Delta \omega i = \omega i - \omega i, ref, \Delta \dot{V}xi = \dot{V}i - \dot{V}xi, ref, \Delta \dot{\omega}i = \dot{\omega}i - \dot{\omega}i, ref \quad (2.2.3.5)$

$$\Delta \lambda i = \begin{bmatrix} \frac{\partial h}{\partial Vxi} & \frac{\partial h}{\partial \omega i} \end{bmatrix} |_{Vxi=Vxi,ref\ \omega i=\omega i,ref} \begin{bmatrix} \Delta Vxi \\ \Delta \omega i \end{bmatrix}, \Delta \lambda i = \lambda i - \lambda i, ref \quad (2.2.3.6)$$

The incremental model therefore is:

$$A11(t) = -gDcos(Ctan^{-1}\left(B\frac{Vxi,ref - \omega i,refR}{Vxi,ref}\right)C\frac{1}{1+\left(B\frac{Vxi,ref - \omega i,refR}{Vxi,ref}\right)^2}B\frac{\omega i,refR}{Vxi,ref^2})$$

$$A12(t) = -gDcos(Ctan^{-1}\left(B\frac{Vxi,ref - \omega i,refR}{Vxi,ref}\right)C\frac{1}{1+\left(B\frac{Vxi,ref - \omega i,refR}{Vxi,ref}\right)^2}B\frac{-R}{Vxi,ref})$$

$$A21(t) = \left(\frac{R}{J}\right)(m \pm \Delta mi,ref)gDcos(Ctan^{-1}\left(B\frac{Vxi,ref - \omega i,refR}{Vxi,ref}\right)C\frac{1}{1+\left(B\frac{Vxi,ref - \omega i,refR}{Vxi,ref}\right)^2}B\frac{\omega i,refR}{Vxi,ref^2})$$

$$A22(t) = \left(\frac{R}{J}\right)(m \pm \Delta mi,ref)gDcos(Ctan^{-1}\left(B\frac{Vxi,ref - \omega i,refR}{Vxi,ref}\right)C\frac{1}{1+\left(B\frac{Vxi,ref - \omega i,refR}{Vxi,ref}\right)^2}B\frac{-R}{Vxi,ref})$$

$$\begin{bmatrix} \Delta \dot{V}xi \\ \Delta \dot{\omega} i \end{bmatrix} = \begin{bmatrix} A11(t) & A12(t) \\ A21(t) & A22(t) \end{bmatrix} \begin{bmatrix} \Delta Vxi \\ \Delta \omega i \end{bmatrix} + \begin{bmatrix} 0 \\ \left(\frac{1}{J}\right) \end{bmatrix} \Delta Ti \quad (2.2.3.7)$$

$$\Delta \lambda i = \begin{bmatrix} \frac{\omega i,refR}{Vxi,ref^2} & \frac{-R}{Vxi,ref} \end{bmatrix} \begin{bmatrix} \Delta Vxi \\ \Delta \omega i \end{bmatrix} \quad (2.2.3.8)$$

Equations *(2.2.3.7), (2.2.3.8)* form a Linear Time Varying (LTV) system. The LTV has the desired state space model format:

$$\Delta \dot{x} = A(t)\Delta x + B(t)\Delta u \quad (2.2.3.9)$$

$$\Delta y = C(t)\Delta x \text{ (22)} \quad (2.2.3.10)$$

The control law that stabilizes the system's trajectory about the reference conditions is defined as follows:

$$u(t) = u, ref(t) + \Delta u^*(t) \quad (2.2.3.11)$$

$$\Delta u^*(t) = -K(x, ref(t))\Delta x(t) \quad (2.2.3.12)$$

The scheduled gain $K(x, ref(t))$ was obtained by solving the Continuous Algebraic Riccati Equation [34]:

$$A(t)^T P(t) + P(t)A(t) - P(t)B(t)R^{-1}B(t)^T P(t) + Q = 0 \quad (2.2.3.13)$$

$$K(x, ref(t)) = R^{-1}B(t)^T P(t) \quad (2.2.3.14)$$

Finally, the positive semi-definite and definite respectively $Q$ and $R$ matrices were designed by prioritizing state error minimization/ convergence against control effort in the LQR optimization problem:

$$minJ(u) = \min (\int_0^{t,final}(\Delta x^T Q \Delta x + \Delta u^T R \Delta u)dt \quad (2.2.3.15)$$

subject to:

$$\Delta \dot{x} = A(t)\Delta x + B(t)\Delta u$$

## 3. Simulation Results

The system was implemented in MATLAB and simulations were performed with MATLAB/SIMULINK. To demonstrate the validity of the control approach for the momentary assisted autonomous emergency braking, we are presenting the following virtual test case corresponding to the following extreme condition: The Lead Vehicle is 10 m ahead of the EGO vehicle in the longitudinal direction (no lateral offset), when it suddenly decelerates with -8 m/s$^2$. Both vehicles were at 100 km/h at that moment, when the Lead Vehicle starts to heavily decelerate. According to *(2.2.1.1.2)*, the minimum braking distance for the EGO vehicle travelling with 100 km/h, on a flat, dry road ($\mu$=0.9), to be brought to a complete halt, plus the additional distance margin, which was set to 1 meter, is approximately 45 meters (Figure 4). This value corresponds to the relative distance threshold $\Delta x, Thres$ as previously discussed. The aforementioned scenario could possibly correspond to an extreme case in which the EGO vehicle was not controlled by any means before, so it is totally to the power of the momentary assisted AEB to mitigate the severity of the collision risk and/or ideally to avoid the collision.

The AEB algorithm managed to safely decelerate EGO vehicle and bring the vehicle to a complete halt respecting the safety distance margin of 1 meters (Figs 4 and 9). The simulation ended when the threat had been repelled and the speed of both vehicles was about 0. Both low level controllers, Sliding Mode controller and Gain-Scheduled LQR managed to provide sufficient deceleration by controlling the slip and thus had successfully followed the desired longitudinal velocity trace, as it can be seen in Figs 3 and 8. However, the Sliding Model controller repelled the collision risk faster than the Linear Quadratic Regulator (Figs 3 and 8). In addition to this, in terms of longitudinal slip error, Sliding Mode achieved near 0% relative error in the first emergency deceleration phase whilst the relative slip error of LQR was 4.4%, respectively (Figs 6, 7 and 11, 12, respectively). The latter could be explained by the fact that LQR was designed about a reference *{wheel translational/rotational speed, braking torque}* trajectory based on an approximation model, which is an estimate of the real system while the SMC was robustly designed in the actual, non-linear vehicle model to reduce slip error to zero.

Results are illustrated in Figures 3-12 where it can be seen that the overall goal has been achieved.

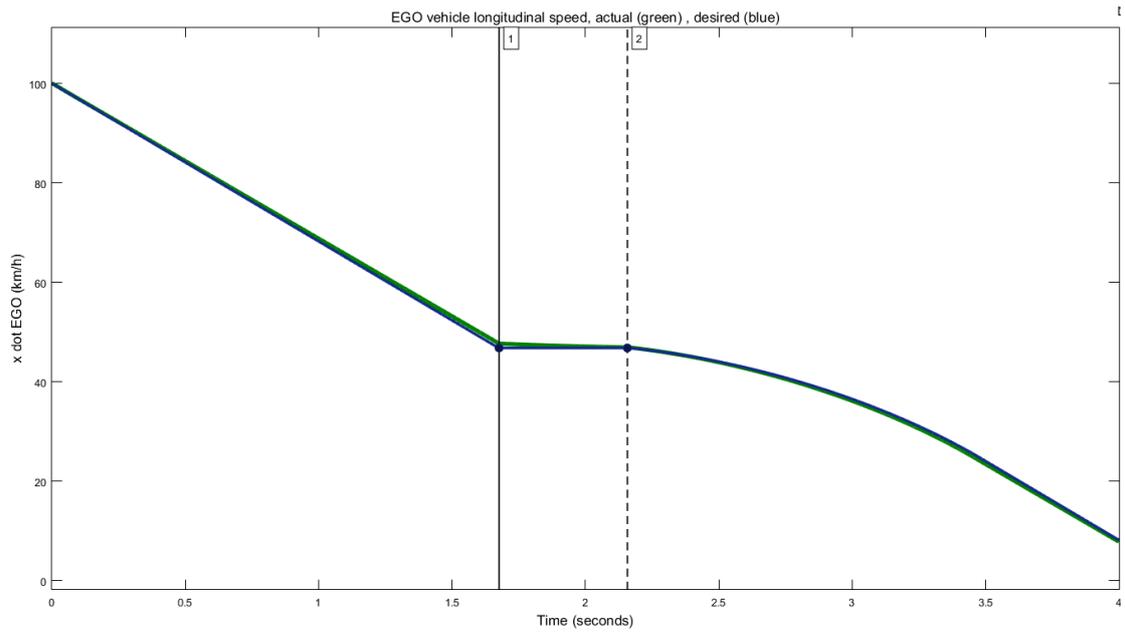

*Figure 3. Illustration of desired longitudinal speed trace(blue line) and actual longitudinal speed trace (green line) of the EGO with RBSC and Sliding Mode Wheel Slip Control, the vertical lines highlighting the time interval (between t1=1.676-t2=2.157 seconds) when the potential collision risk is repelled, and the speed regulator is on*

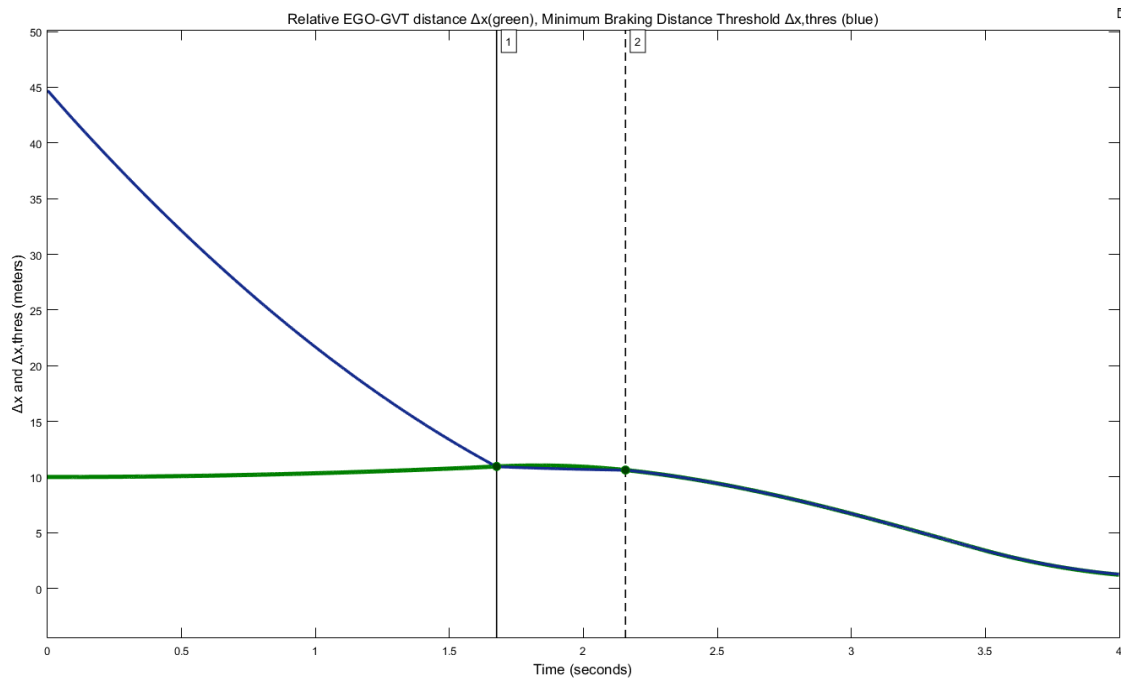

*Figure 4. Illustration of the relative EGO- GVT vehicle distance (green line) and relative distance threshold (blue line). RBSC and Sliding Mode Wheel Slip Control adapts to the relative distance threshold in less than 1.7 sec and then retains the required distance limit.*

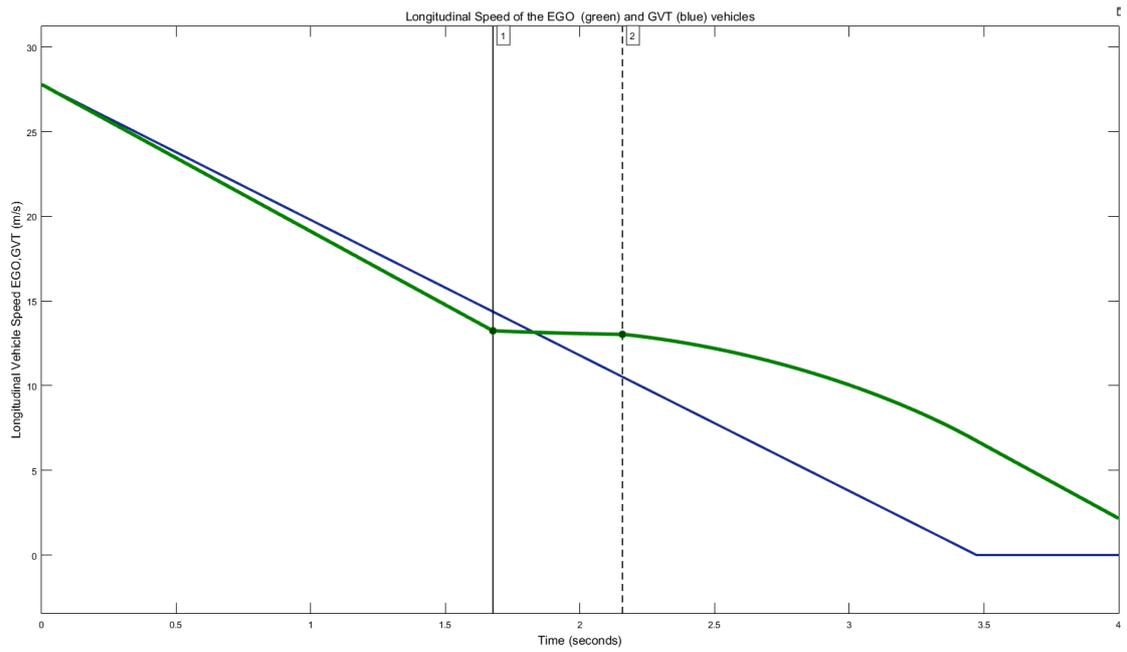

*Figure 5. Illustration of the EGO vehicle (equipped with RBSC and SMC) longitudinal speed (green line) and Lead vehicle (GVT) longitudinal speed (blue line)*

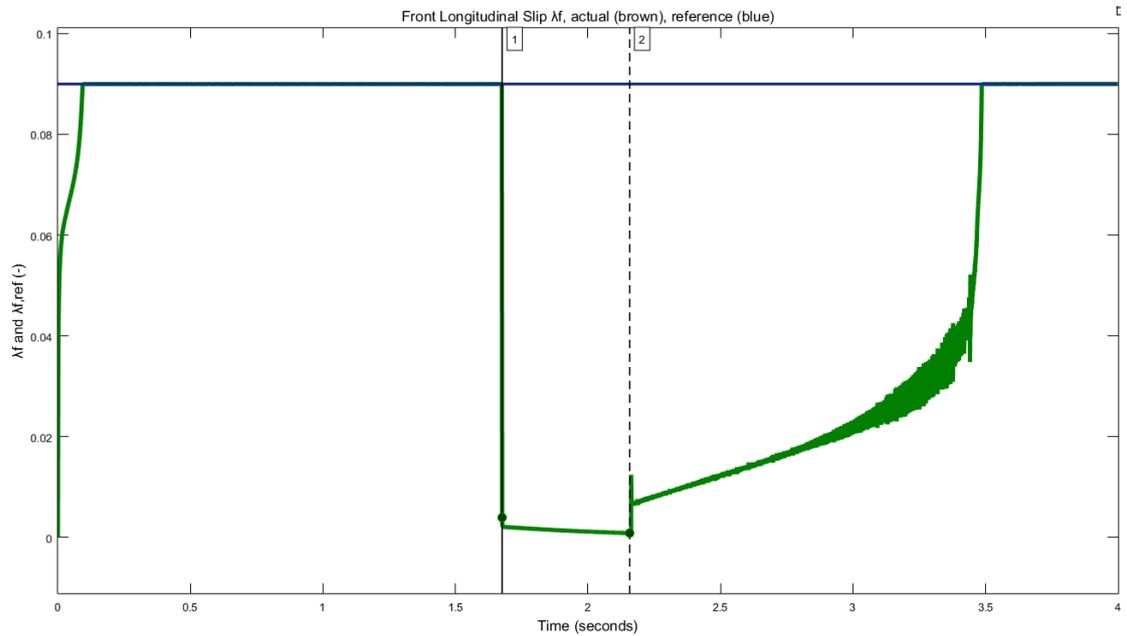

*Figure 6. Illustration of the Sliding Mode Wheel Slip Controller performance regarding front wheel slip target, whereas it can be seen that between t1=1.676-t2=2.157 seconds wheel slip control is deactivated whilst PID Speed Regulator is aiming at maintaining a cruising speed.*

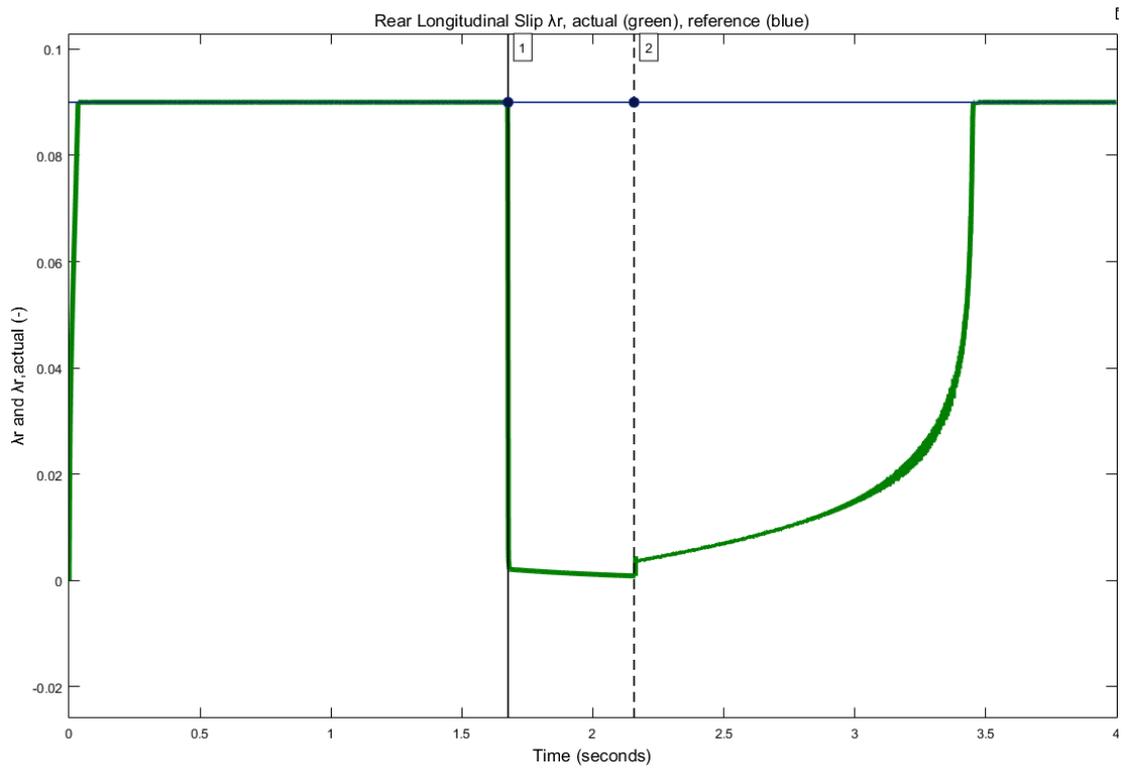

*Figure 7. Illustration of the Sliding Mode Wheel Slip Controller performance regarding rear wheel slip target, whereas it can be seen that between t1=1.676-t2=2.157 seconds wheel slip control is deactivated whilst PID Speed Regulator is aiming at maintaining a cruising speed.*

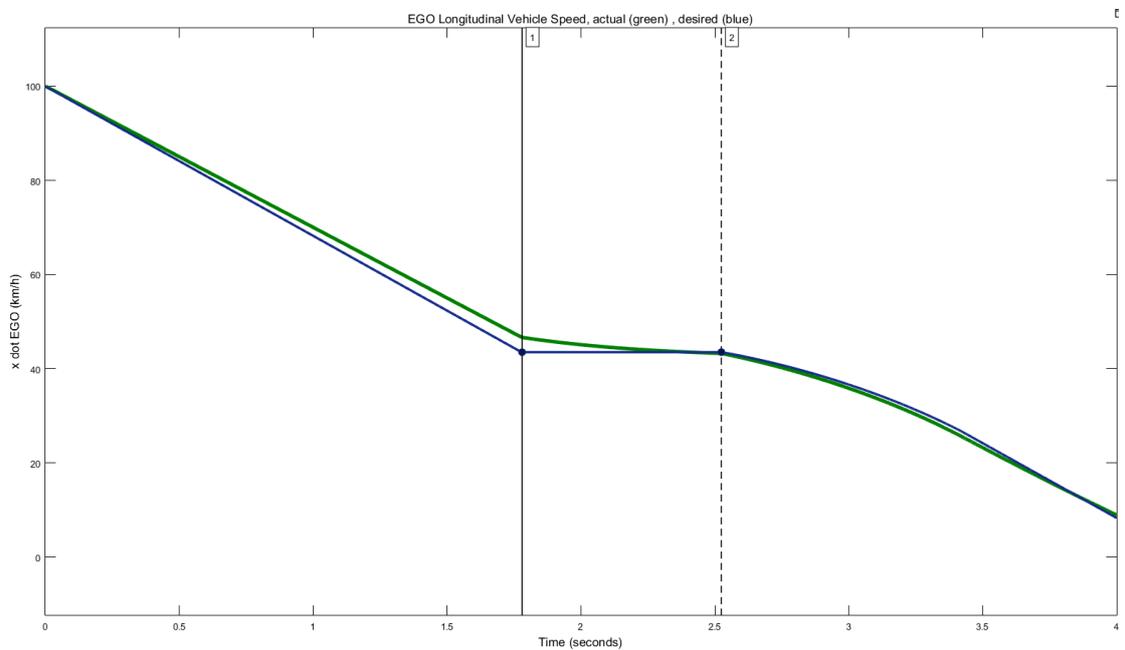

*Figure 8. Illustration of desired longitudinal speed trace(blue line) and actual longitudinal speed trace (green line) of the EGO with RBSC and LQR, the vertical lines highlighting the time interval (between t1=1.780-t2=2.524 seconds) when the potential collision risk is repelled, and the speed regulator is on*

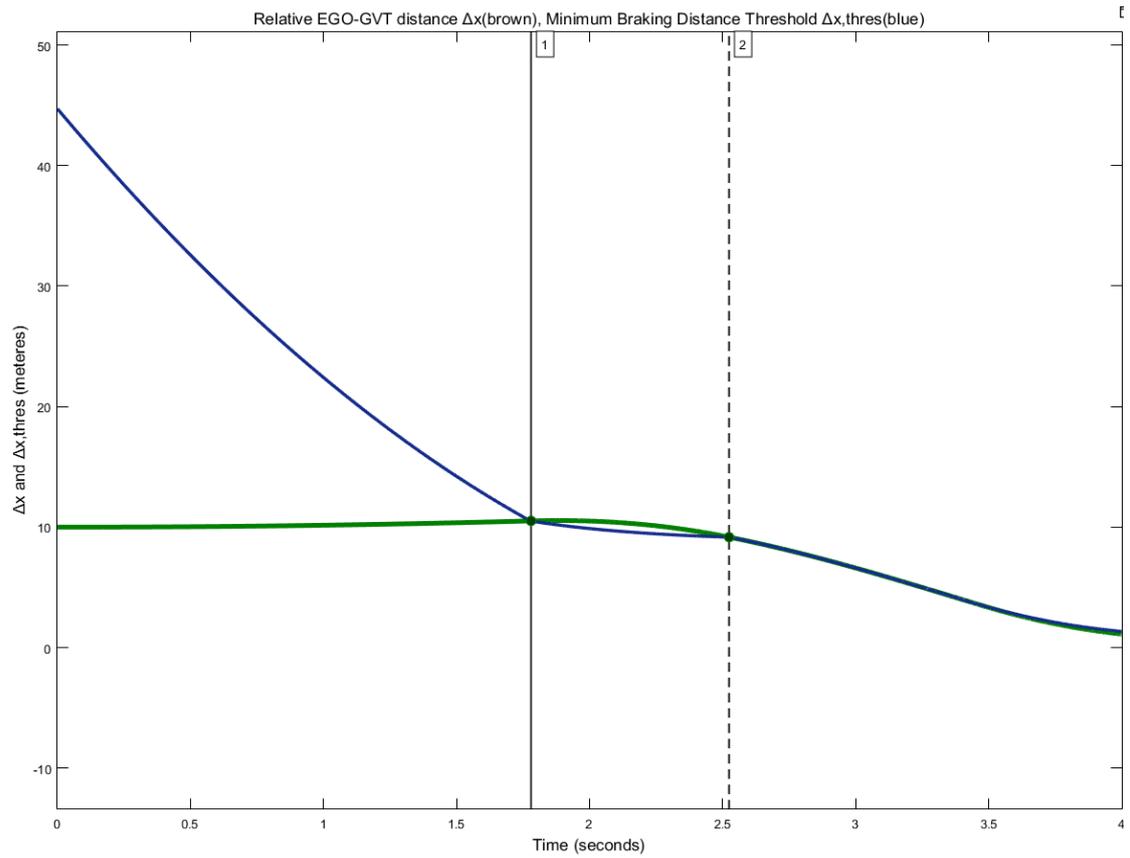

*Figure 9. Illustration of the relative EGO- GVT vehicle distance (green line) and relative distance threshold (blue line). RBSC and LQR sufficiently adapts to the relative distance threshold (in less than 1.8 sec) and then retains the required distance limit.*

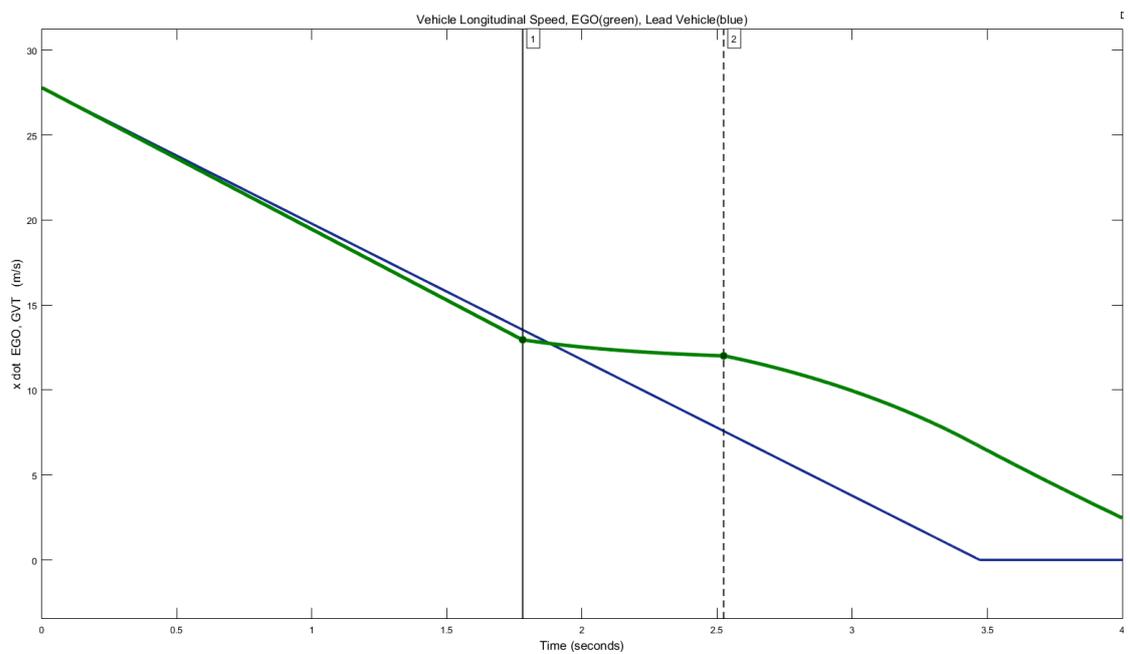

*Figure 10. Illustration of the EGO vehicle (equipped with RBSC and Gain-Scheduled LQR) longitudinal speed (green line) and Lead vehicle (GVT) longitudinal speed (blue line)*

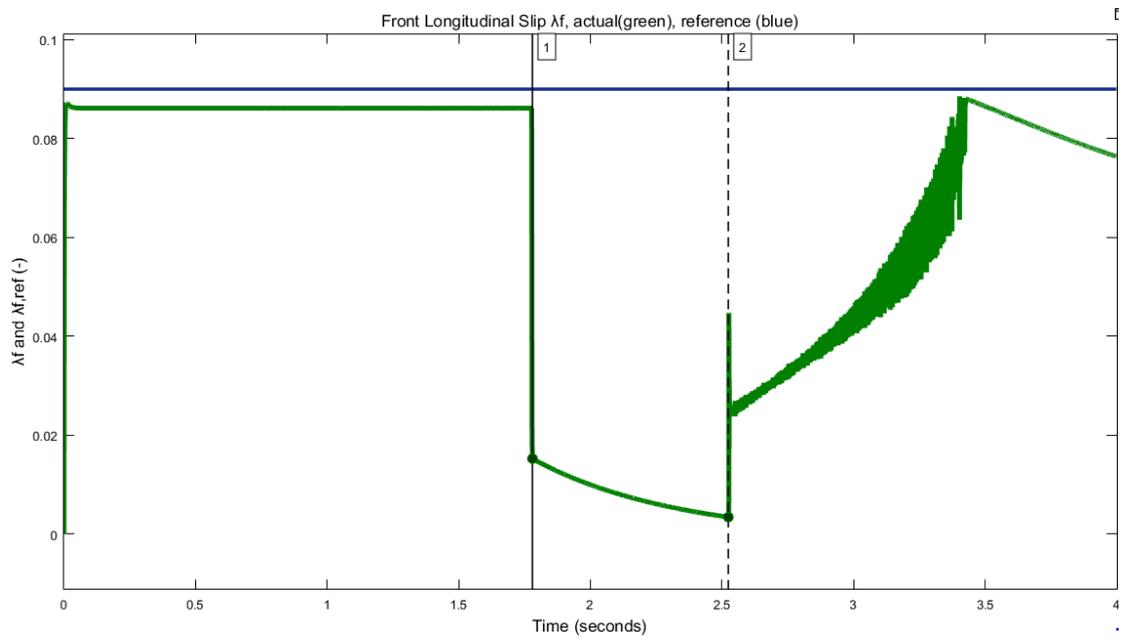

*Figure 11. Illustration of the LQR slip controller performance regarding front wheel slip target .It can be seen that between t1=1.780 and t2=2.524 seconds, wheel slip control is deactivated whilst PID Speed Regulator is aiming at maintaining a cruising speed.*

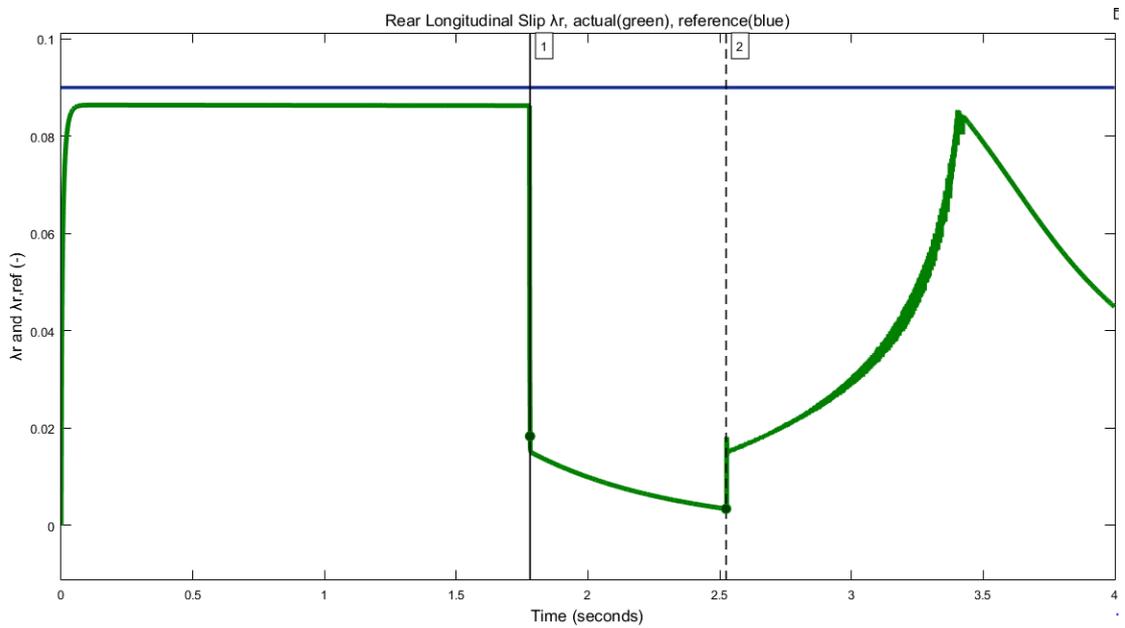

*Figure 12. Figure 3.9 Illustration of the LQR slip controller performance regarding rear wheel slip target. It can be seen that between t1=1.780 and t2=2.524 seconds, wheel slip control is deactivated whilst PID Speed Regulator is aiming at maintaining a cruising speed.*

## 4. Conclusions

In this study, a safety distance based hierarchical AEB control system was proposed. The hierarchical AEB control structure is constituted of a) a high-level Rule-Based Supervisory control module, b) an intermediate-level switching algorithm, and c) a low-level control module. The control system was also augmented with a Speed Regulator for the in-between collision threat phases that the EGO vehicle undergoes. Two distinct control design approaches were studied, differing only in the low level. The first incorporated a Robust Sliding Mode wheel slip control and the second a Gain-Scheduled Linear Quadratic Regulator for heavy deceleration trajectory stabilization. The two control system combinations were validated in Simulink through a straight-line emergency braking maneuver simulation.

During the emergency deceleration phase, the AEB system with Sliding Mode low-level control achieved longitudinal slip relative error near 0%, whilst the overall speed trace following ability was satisfactory and the defined safety distance threshold was respected. At the end of the simulated emergency maneuver, the desired relative distance of 1 meter with respect to the leading vehicle was successfully achieved and maintained. The AEB system with LQR low-level control achieved a greater longitudinal slip relative error compared to Sliding Mode controller, accounting for approximately 4.4% during the emergency deceleration phase. The overall speed trace following ability was satisfactory and the defined safety distance threshold was respected. At the end of the simulated emergency maneuver, the desired relative distance of 1 meter with respect to the leading vehicle was successfully achieved and maintained, as in the case of Sliding Mode low-level control. Therefore, full collision avoidance was achieved in both proposed control system combinations. The hierarchical control structure proposed in this study is flexible and extendable to be implemented on autonomous vehicles for momentary assisted emergency braking.